\documentstyle[12pt,epsfig]{article}
 
\begin{document}                                                                     
\baselineskip 22pt                                                                   
\noindent
\hspace*{11.5cm}
SNUTP 95-065\\
\noindent
\hspace*{11.7cm}
YUMS 95-017\\
\noindent
\hspace*{12.0cm}
(June 1995)\\

\begin{center}                                                                                                                                  
{\Large \bf Decay Constants and Semileptonic Decays\\
of Heavy Mesons in Relativistic Quark Model
\\}
                                                    
\vspace{0.2cm}                                                                                                                                  
Dae Sung Hwang$^1$, ~C. S. Kim$^2$ ~and~ Wuk Namgung$^3$\\

$1$: Department of Physics, Sejong University, Seoul 133--747, Korea\\
$2$: Department of Physics, Yonsei University, Seoul 120--749, Korea\\
$3$: Department of Physics, Dongguk University, Seoul 100--715,
Korea\\
																				
\vspace{0.5cm}                                                                                                                                  
																				
{\bf Abstract} \\
\end{center}                                                                                                                                    

We investigate the $B$ and $D$ mesons in the relativistic quark model
by applying the variational method with the Gaussian wave function.
We calculate the Fermi momentum parameter $p_{_F}$,
and obtain $p_{_F} = 0.50 \sim 0.54$ GeV, which is almost independent
of the input parameters, $\alpha_s$, $m_b$, $m_c$ and $m_{sp}$.
We then calculate the ratio $f_B$/$f_D$, 
and obtain the result which is larger, by 
the factor of about 1.3, than $\sqrt{M_D / M_B}$ given by the naive 
nonrelativistic analogy.
This result is in a good agreement with the recent Lattice
calculations.
We also calculate the ratio $(M_{B^*}-M_{B})$/$(M_{D^*}-M_{D})$.
In these calculations the wave function at origin $\psi (0)$ is
essential.
We also determine $p_{_F}$ by comparing
the theoretical prediction
of the ACCMM model with the lepton energy spectrum of
$B \rightarrow e \nu X$ from the recent ARGUS analysis, and find that
$p_{_F}=0.27~\pm~^{0.22}_{0.27}$ GeV, when we use $m_c=1.5$ GeV.
However, this experimentally determined value of $p_{_F}$ is strongly
dependent on the value of input parameter $m_c$.

\vspace{1.0cm}
                                                                                
\vfill

\noindent
$1$: e-mail: dshwang@phy.sejong.ac.kr\\
$2$: e-mail: kim@cskim.yonsei.ac.kr
\thispagestyle{empty}                                                           
\pagebreak                                                                                                                                              

\baselineskip 22pt

\noindent
{\bf \large 1. Introduction}\\

The $B$ meson physics is important in the present high energy physics,
since it gives us the information on the CKM matrix elements,
$V_{cb}$ and $V_{ub}$,
and it is expected to provide the CP violation phenomena.
In order to extract the value of $V_{ub}$ from the $B$ meson decay
experiments, the method of separating the $B\rightarrow X_u l\nu$
events from the $B\rightarrow X_c l\nu$ ones at the end point region
of the lepton energy spectrum has been used \cite{cleo,argus}.
On the other hand, the method of using the hadronic
invariant mass spectrum has also been suggested recently \cite{kim}.
For the analysis of the inclusive semileptonic decay process, 
the ACCMM model \cite{alta} has been most popularly imployed,
where the Fermi momentum parameter $p_{_F}$ is introduced as the most 
important parameter. The value of $p_{_F} \simeq 0.3$ GeV has been 
commonly used for the experimental analyses without
clear experimental or theoretical supports. 
In reference \cite{hkn} we calculated $p_{_F}$ in the relativistic
quark model using the variational method with the Gaussian wave
function, and obtained the result of $p_{_F}$=0.54 GeV.
In this paper we also study
the ratios, $f_B$/$f_D$ and $(M_{B^*}-M_{B})$/$(M_{D^*}-M_{D})$,
using the same method.

When one treats the heavy-light meson in analogy with the
nonrelativistic situation, one expects
$f_B / f_D \simeq {\sqrt{M_D / M_B}}$, since the reduced masses
of the light and heavy quark systems of the $B$ and $D$ mesons have
similar values, and $f_P^{\ 2}M_P=12\, |\psi (0)|^2$ 
by the van Royen--Weisskopf formula
for the pseudoscalar meson $P$ \cite{royen}, where $\psi (0)$
is the wave function at the origin of the relative motion of quarks.
If one uses the relation $f_B / f_D \simeq {\sqrt{M_D / M_B}}$ 
with the supplementary relation
$f_D / f_{D_s}\simeq {\sqrt{m_d / m_s}}$,
one can obtain the value of $f_B$ from that of $f_{D_s}$, which
is rather well established from the branching ratio of the
$\bar B^0 \rightarrow D^+ D_s^-$ or
$D_s^+\rightarrow {\mu}^+ \nu_{\mu}$ process \cite{rosner}.
However, our calculation of $f_P$ in the relativistic quark
model, which we present in this article, shows that the above
consideration with nonrelativistic analogy is much deviated by
the relativistic motion of the light quark in the heavy-light
pseudoscalar meson $P$.
This deviation has also been exposed by the recent Lattice
calculations, since they give rather close values for 
$f_B$ and $f_D$ \cite{latt,bern}.
This situation can be understood clearly within our relativistic
calculation.

The potential model has been successful for $\psi$ and $\Upsilon$
families with the nonrelativistic Hamiltonian, since their heavy
quarks can be treated nonrelativistically.
However, for $D$ or $B$ meson it has been difficult to apply
the potential model because of the relativistic motion of the
light quark in $D$ or $B$ meson.
In our calculation we work with the realistic Hamiltonian
which is relativistic for the 
light quark and nonrelativistic for the heavy quark,
and adopt the variational method.
We take the Gaussian function as the trial wave function, and obtain
the ground state energy (and the wave function) by minimizing the 
expectation value of the Hamiltonian. 
Using the Gaussian wave function calculated as above, 
we get the wave function at the origin $\psi (0)$, and
with which we can calculate the decay constant 
of the heavy-light pseudoscalar meson from the van Royen-Weisskopf
formula. 
Through this procedure we obtain the ratio $f_B$/$f_D$.
We also calculate the ratio $(M_{B^*}-M_{B})$/$(M_{D^*}-M_{D})$
from the chromomagnetic hyperfine splitting formula,
where the information of $\psi (0)$ is  essential.
Finally, we compare the value of the Fermi momentum $p_{_F}$ given
in our calculation with the lepton energy spectrum data of the
semileptonic decay process,
and find that it is just outside off one $\sigma$ standard deviation.

In Section 2, using variational method in relativistic quark model 
we calculate the value of the parameter $p_{_F}$, and the ratios  
$f_B/f_D$ and $(M_{B^*}-M_B)/(M_{D^*}-M_D)$. We also extract $p_{_F}$
by comparing the theoretical prediction of the ACCMM model with the
whole region of electron energy spectrum of $B \rightarrow e \nu X$
in Section 3.
Section 4 contains the conclusions.
\\

\pagebreak

\noindent
{\bf \large 2. Variational method in relativistic quark model, and 
calculations\\
\hspace*{0.5cm}
of $f_B$/$f_D$ and $(M_{B^*}-M_{B})$/$(M_{D^*}-M_{D})$}\\

For the $B$ meson system we treat the $b$-quark nonrelativistically,
but the $u$- or $d$-quark relativistically with the Hamiltonian
\begin{equation}
H=M+{{{\bf p}^2}\over {2M}}+{\sqrt{{\bf p}^2+m^2}}+V(r),
\label{f8}
\end{equation}
where $M=m_b$ or $m_c$ is the heavy quark mass and 
$m=m_{sp}$ is the $u$- or $d$-quark mass, {\it i.e.}
the spectator quark mass in the ACCMM model.
We apply the variational method to the Hamiltonian (\ref{f8})
with the trial wave function
\begin{equation}
\psi ({\bf r})=({{\mu}\over {\sqrt{\pi}}})^{3/2}e^{-{\mu}^2r^2/2},
\label{f9}
\end{equation}
where $\mu$ is the variational parameter.
The Fourier transform of $\psi ({\bf r})$ gives the momentum space
wave function $\chi ({\bf p})$, which is also Gaussian,
\begin{equation}
\chi ({\bf p}) = {{1}\over {({\sqrt{\pi}}{\mu})^{3/2}}}
e^{-p^2/2{\mu}^2}.
\label{f5}
\end{equation}
The ground state is given by minimizing the expectation value of $H$,
\begin{equation}
\langle H\rangle =\langle\psi\vert H\vert\psi\rangle =E(\mu ),
\ \ \
{{d}\over {d\mu }}E(\mu )=0\ \ {\rm{at}}\ \ \mu ={\bar{\mu}},
\label{f10}
\end{equation}
and then $\bar E \equiv E({\bar{\mu}})$ approximates $m_B$, and we
get ${\bar{\mu}} = p_{_F}$, the Fermi momentum parameter in the ACCMM
model.
The value of $\mu$ or $p_{_F}$ corresponds to the
measure of the radius of the two body bound state, as can be seen from
the relation,
$\langle r\rangle ={{2}\over{\sqrt{\pi}}}{{1}\over \mu }$
or $\langle r^2{\rangle}^{{1}\over {2}} ={{3}\over {2}}
{{1}\over \mu }$.

In Eq. (\ref{f8}), we take the Cornell potential
which is composed of the Coulomb and linear potentials,
\begin{equation}
V(r)=-{{{\alpha}_c}\over {r}}+Kr.
\label{f13}
\end{equation}
For the values of the parameters
${\alpha}_c\ (\equiv {{4}\over {3}}{\alpha}_s)$, 
the quark masses $m_b$ and $m_c$, and $K$, we use the following
two sets of parameters.
The set $\{\rm{A}\}$ is
the Hagiwara $et$ $al.$'s~\cite{hagi} which has been determined by 
the best fit of $(c{\bar{c}})$ and $(b{\bar{b}})$ bound state spectra.
And the set $\{\rm{B}\}$ is chosen to have the running coupling
constants at the mass scales of $m_B$ and $m_D$, and the quark masses
$m_b$ and $m_c$ which were determined to give the best $\psi$ and
$\Upsilon$ masses for the variational ground states.
\begin{eqnarray}
\{ {\rm{A}} \} ~~~~~
{\alpha}_c=0.47,\ m_b=4.75\ GeV,\ m_c=1.32\ GeV,\ K=0.19\ GeV^2,
\label{f14a}\\ 
\{ {\rm{B}} \} 
{\alpha}_c^{\, B}=0.32,\ m_b=4.64\ GeV;\ {\alpha}_c^{\, D}=0.48,
\ m_c=1.33\ GeV;\ K=0.19\ GeV^2.
\label{f14b}
\end{eqnarray}

With the Gaussian trial wave function (\ref{f9}) or (\ref{f5}),
the expectation value of each term of
the Hamiltonian (\ref{f8}) is given as follows:
\begin{eqnarray}
\langle {{\bf p}^2\over 2M}\rangle &=& 
\langle \chi ({\bf p}\,) | {{\bf p}^2\over 2M}|
\chi ({\bf p}\,) \rangle = {3 \over 4M} \mu^2,
\nonumber\\
\langle \sqrt{{\bf p}\,^2+ m^2} \rangle &=& \langle \chi ({\bf p}\,) 
| \sqrt{{\bf p}\,^2+ m^2} | \chi ({\bf p}\,) \rangle 
= {4\mu \over \sqrt\pi} \int_0^\infty e^{-x^2} \sqrt{x^2 + (m/\mu)^2}
\; x^2dx,
\nonumber\\
\langle V(r) \rangle &=& \langle \psi ({\bf r}) | -{\alpha_c \over r}
+ Kr \
|\psi ({\bf r}) \rangle 
= {2 \over \sqrt\pi} (-\alpha_c\mu + {K / \mu} ).
\label{f22}
\end{eqnarray}
Then we have
\begin{eqnarray}
E({\mu})&=&\langle H\rangle
\label{f21s}\\
&=& M+
{3 \over 4M} \mu^2 +
{2 \over \sqrt\pi} (-\alpha_c\mu + {K / \mu} ) +
{4\mu \over \sqrt\pi} \int_0^\infty e^{-x^2} \sqrt{x^2 + (m/\mu)^2}
\; x^2dx.
\nonumber
\end{eqnarray}
For more details on this procedure of the variational method, 
see Ref. \cite{hkn}.

With  the input value of $m =m_{sp} = 0.15$ GeV,
which is the value commonly used in experimental analyses,
we minimize
$E({\mu})$ of (\ref{f21s}), and then we obtain for the $B$ meson,
\begin{eqnarray}
p_{_F}(B)=\bar \mu &=& 0.54 \ {\rm GeV}, \qquad \bar{E}(B) = 5.54 \ 
{\rm GeV} 
\qquad {\rm for\ \ \{ A \} },
\label{f31}\\
\bar \mu &=& 0.50 \ {\rm GeV}, \qquad \bar{E}(B) = 5.52 \ {\rm GeV} 
\qquad {\rm for\ \ \{ B \} }.
\nonumber
\end{eqnarray}
The $B$ meson mass is lowered from the above values
if we include chromomagnetic hyperfine splitting corrections.
For comparison, let us check how much sensitive our calculation of
$p_{_F}$ is by considering the case where $m=m_{sp}=0$.
For $m_{sp}=0$ the integral in (\ref{f21s}) is done easily and we
obtain the following values of $\bar \mu =p_{_F}$, 
\begin{eqnarray}
\bar \mu &=& 0.53 \ {\rm GeV}, \qquad\bar{E}(B) = 5.52 \ {\rm GeV}
\qquad {\rm for\ \ \{ A \} },
\label{f32}\\
\bar \mu &=& 0.48 \ {\rm GeV}, \qquad\bar{E}(B) = 5.49 \ {\rm GeV}
\qquad {\rm for\ \ \{ B \}}.
\nonumber
\end{eqnarray}
As we see in Eq. (\ref{f32}), the results are similar to those in 
(\ref{f31}), where $m_{sp}=0.15$ GeV.
We expected this insensitivity of the value of $p_{_F}$ on
$m_{sp}$ because the value of $m_{sp}$, which should be small
in any case, can not affect the integral in (\ref{f21s})
significantly.
We also note that the theoretically determined value of $p_{_F}(B)$ 
is completely independent of the input value of $m_c$, as can be seen 
from Eq. (\ref{f21s}). Following the same procedure,
we next obtain the results for the $D$ meson with
$m_{sp}=0.15$ GeV,
\begin{eqnarray}
p_{_F}(D)=\bar \mu &=& 0.45 \ {\rm GeV}, \qquad \bar{E}(D) = 2.21 \ 
{\rm GeV} 
\qquad {\rm for\ \ \{ A \} },
\label{f31a}\\
\bar \mu &=& 0.46 \ {\rm GeV}, \qquad \bar{E}(D) = 2.21 \ 
{\rm GeV}  \qquad
{\rm for\ \ \{ B \} }.
\nonumber
\end{eqnarray}

The decay constant $f_P$ of a pseudoscalar meson $P$ is defined by
the matrix element $\langle 0 | A_{\mu} | P(q)\rangle$:
\begin{equation}
\langle 0 | A_{\mu} | P(q)\rangle = i q_{\mu} f_P .
\label{r1}
\end{equation}
By considering the low energy limit of the heavy meson annihilation,
we have the relation between $f_P$ and the ground state wave function
at the origin ${\psi}_P(0)$ from the van Royen-Weisskopf formula
including the color factor \cite{royen,rosner}:
\begin{equation}
f_P^2={12\over M_P}|{\psi}_P(0)|^2,
\label{r2}
\end{equation}
where $M_P$ is the heavy meson mass.
{}From (\ref{r2}) we have the ratio of $f_B$ and $f_D$:
\begin{equation}
{f_B\over f_D}=
\sqrt{{M_D\over M_B}}\times {|{\psi}_B(0)|\over |{\psi}_D(0)|}.
\label{r3}
\end{equation}
For the Gaussian wave function of Eq. (\ref{f9}), we have
\begin{equation}
{\psi}_P(0)=({{p_{_F}(P)}\over {\sqrt{\pi}}})^{3/2},
\label{r3a}
\end{equation}
then using the values of $p_{_F}$ in (\ref{f31}) and (\ref{f31a}),
we obtain
\begin{eqnarray}
{f_B\over f_D}=
\sqrt{{M_D\over M_B}}\times
\Bigl( {p_{_F}(B)\over p_{_F}(D)}\Bigr)^{3/2}
&=&
0.59\times 1.31 =0.77
\qquad {\rm for\ \ \{ A \} },
\label{r4}\\
&=&
0.59\times 1.13 =0.67
\qquad {\rm for\ \ \{ B \} }.
\nonumber
\end{eqnarray}
{}From (\ref{r4}) we see that $f_B / f_D$ is enhanced, compared with
$\sqrt{M_D / M_B}$, by the factor 1.31
for the parameter set \{A\}, and by 1.13 for the set \{B\},
which are given by the factor of $|\psi_B(0)/\psi_D(0)|$.
Sometimes this factor has been approximated to be 1, and
the relation $f_B/f_D\simeq \sqrt{M_D / M_B}$ has been used,
by treating it in analogy with the nonrelativistic case \cite{rosner}.
However, our calculation shows that this factor is indeed important
and different from 1 significantly.
The factor 1.31 obtained in (\ref{r4}) for the parameter set \{A\} is
in a pretty good agreement with 1.40 of Ref. \cite{latt} and 1.39
of Ref. \cite{bern} of the recent Lattice calculations.

The mass difference between the vector meson $P^*$ and
the pseudoscalar meson $P$ is given
rise to by the
chromomagnetic hyperfine splitting:
\begin{equation}
V_{hf} = {2 \over 3 m_Q{\widetilde{m}}_q} \;
\vec s_1 \cdot \vec s_2 \,\nabla^2
(- {\alpha_c \over r}),
\label{f33}
\end{equation}
where $m_Q$ is the heavy quark mass as we used before 
({\it i.e.} $m_b$ or $m_c$), and ${\widetilde{m}}_q$ is
the constituent quark mass of the light quark, which is the effective
mass for the baryon magnetic moments \cite{lich},
\begin{equation}
{\widetilde{m}}_u={\widetilde{m}}_d=0.33\ {\rm GeV},\ \
{\widetilde{m}}_s=0.53\ {\rm GeV}.
\label{f33s}
\end{equation}
Then the mass difference between $B^*$ and $B$ mesons is given by
\begin{equation}
M_{B^*}-M_{B}=
{8\pi \alpha_c \over 3 m_b {\widetilde{m}}_u}|{\psi}_B (0)|^2.
\label{f34}
\end{equation}
Using the values of $p_{_F}$ in (\ref{f31}) and (\ref{f31a}),
we obtain the ratio of $M_{B^*}-M_{B}$ and $M_{D^*}-M_{D}$ as
\begin{eqnarray}
{M_{B^*}-M_{B}\over M_{D^*}-M_{D}}
= {m_c \over m_b} &\times&
\Bigl({|{\psi}_B(0)|\over |{\psi}_D(0)|}\Bigr)^2
={m_c \over m_b} \times
\Bigl( {p_{_F}(B)\over p_{_F}(D)}\Bigr)^3
\nonumber\\
&=& 0.28 \times 1.73=0.48
\qquad {\rm for\ \ \{ A \} },
\label{f34a}\\
&=& 0.29 \times 1.28=0.37
\qquad {\rm for\ \ \{ B \} }.
\nonumber
\end{eqnarray}
The experimental value \cite{pdg} is about 0.33, which is larger than
the nonrelativistic value 0.28 or 0.29, but smaller than our
calculated value 0.48 or 0.37.
Our calculated results are at least not much 
worse than the nonrelativistic values, even though there exists
somewhat large discrepancy compared to experimental result.
And this suggests that the behind reason for this discrepancy 
could be more subtle than the nonrelativistic consideration.
\\

\noindent
{\bf \large 3. Determination  of $p_{_F}$ from the experimental
spectrum}\\

Until now, we have discussed the theoretical determination of $p_{_F}$
in the relativistic quark model using the variational method, and its
implications to the heavy meson masses and the decay constants of
the heavy mesons. Now we would like to determine the Fermi momentum
parameter $p_{_F}$ by comparing the theoretical prediction 
with the experimental
charged lepton energy spectrum in semileptonic decays of $B$ meson.

As discussed, the simplest model for the semileptonic $B$-decay is 
the spectator model which 
considers the decaying $b$-quark in the $B$ meson as a free particle. 
The spectator model is usually used with the inclusion of perturbative
QCD radiative corrections \cite{kuhn}.
Then the decay width of the process $B\rightarrow X_ql\nu$ is given by
\begin{equation}
{\Gamma}_B (B\rightarrow X_ql\nu )\simeq
{\Gamma}_b (b\rightarrow ql\nu )=
\vert V_{bq}{\vert}^2
({{G_F^2m_b^5}\over {192{\pi}^3}})f({{m_q}\over {m_b}})
[1-{{2}\over {3}}
{{{\alpha}_s}\over {\pi}}g({{m_q}\over {m_b}})],
\label{f1}
\end{equation}
where $m_q$ is the mass of the final $q$-quark decayed from $b$-quark.
As can be seen, the decay width of the spectator model depends on
$m_b^5$, 
therefore small difference of $m_b$ would change the decay width
significantly.

Altarelli $et$ $al.$ \cite{alta} proposed for the inclusive $B$ meson 
semileptonic decays their ACCMM model, which incorporates the bound
state effect by treating the $b$-quark as a virtual state particle,
thus giving momentum dependence to the $b$-quark mass. 
The virtual state $b$-quark mass 
$W$ is given by
\begin{equation}
W^2({\bf p})=m_B^2+m_{sp}^2-2m_B{\sqrt{{\bf p}^2+m_{sp}^2}}
\label{f2}
\end{equation}
in the $B$ meson rest frame, where $m_B$ is the $B$ meson mass,
$m_{sp}$ is the spectator quark mass, and {\bf p} is the momentum of 
the spectator quark inside $B$ meson.

For the momentum distribution of the virtual $b$-quark, Altarelli
$et$ $al.$ considered the Fermi motion inside the $B$ meson with the
Gaussian momentum distribution,
\begin{equation}
\phi ({\bf p}; p_{_F})= 4 \pi | \chi({\bf p}) |^2 = 
{{4}\over {{\sqrt{\pi}}p_{_F}^3}}e^{-{\bf p}^2/p_{_F}^2},
\label{f3}
\end{equation}
where the Gaussian width, $p_{_F}$, is treated as a free parameter.
Then the lepton energy spectrum of the $B$ meson  decay is given by
\begin{equation}
{{d{\Gamma}_B}\over {dE_l}}(p_{_F}, m_{sp}, m_q, m_B)=
{\int}_0^{p_{max}}dp\ p^2\phi ({\bf p}; p_{_F})\ 
{{d{\Gamma}_b}\over{dE_l}}(m_b=W, m_q, m_{sp}),
\label{f4}
\end{equation}
where $p_{max}$ is the maximum kinematically allowed value of
$p=|{\bf p}|$.
The ACCMM model, therefore, introduces a new parameter $p_{_F}$ for
the Gaussian momentum distribution of the $b$-quark inside $B$ meson,
instead of the $b$-quark mass of the spectator model.
In this way the ACCMM model incorporates the bound state effects and
reduces the strong dependence on the $b$-quark mass in the decay width
of the spectator model.
The Fermi momentum parameter $p_{_F}$ is the most essential parameter
of the ACCMM model, as we see in the above.
However, the experimental determination of its value 
from the lepton energy spectrum has been very ambiguous, because
various parameters of ACCMM model, such as $p_{_F}$, $m_q$ and
$m_{sp}$, are fitted all together from the limited region of end-point
lepton energy spectrum,
and because the perturbative QCD corrections are very sensitive
in the end-point region of the spectrum. 

Recently, ARGUS \cite{argus2}
extracted the model independent lepton energy spectrum of 
$B \rightarrow X_c l \nu$ for the whole region of electron energy,
but with much larger uncertainties, as shown in Fig. 1.
We now compare the whole region of experimental electron energy
spectrum with the theoretical prediction of ACCMM model,
Eq. (\ref{f4}), 
using $p_{_F}$ as a free parameter. We fixed $m_{sp}=0.15$ GeV and
$m_q=m_c=1.5$ GeV,  which are the values commonly
used by experimental analyses.
We derive the value of $p_{_F}$ using $\chi^2$ analysis, and we obtain
\begin{equation}
p_{_F}~ =~ 0.27~ \pm~ ^{0.22}_{0.27}~~{\rm GeV}.
\label{z2}
\end{equation}
The minimum $\chi^2$  equals to 0.59 with $p_{_F} = 0.27$ GeV. 
However, the result, Eq. (\ref{z2}),
is found to be strongly dependent on the input value of $m_c$:
if we instead use smaller $m_c$, both the best fit value of $p_{_F}$ 
and the minimum $\chi^2$ increase, and {\it vise versa}.
In Fig. 1, we also show the theoretical ACCMM model spectrums with
$p_{_F}=0,~0.27,~0.49$ GeV, corresponding to dashed-, full-,
dotted-line, respectively.
The experimental data and the theoretical predictions are all 
normalized to the semileptonic branching
ratio, $BR(B \rightarrow e \nu X) = 9.6 \%$,
following the result of ARGUS \cite{argus2}.

In Section 2, we theoretically calculated the Fermi momentum 
parameter $p_{_F}$, and obtained $p_{_F}=0.50 \sim 0.54$ GeV.
We note that the theoretically calculated values are slightly 
outside off one $\sigma$ standard deviation comparing with the best 
fit value of the experimental data.
However, since the experimental spectrum
has still large uncertainties, we cannot exclude the validity of 
relativistic quark model for calculation of $p_{_F}$ yet, nor 
ACCMM model itself to apply to  experimental analyses 
to find CKM parameters, $|V_{cb}|$ and/or $|V_{ub}/V_{cb}|$.
In near future,
once we get much more data from asymmetric $B$ factories,
it would be very interesting to extract the precise value of $p_{_F}$
once again.

In our previous work \cite{hkn}, we investigated the dependence of 
$|V_{ub}/V_{cb}|$ on $p_{_F}$ in ACCMM model, and found rather strong 
dependence as a function of a parameter $p_{_F}$,
\begin{eqnarray}
10^2 \times |V_{ub}/V_{cb}|^2
&=&0.57\pm 0.11 ~~({\rm ACCMM~with}~p_{_F}=0.3~ \cite{cleo2}),
\nonumber\\
&=&1.03\pm 0.11 ~~({\rm ACCMM~with}~p_{_F}=0.5),
\nonumber\\
&=&1.02\pm 0.20 ~~({\rm Isgur}~et~al.~ \cite{isgw}).
\label{g3}
\end{eqnarray} 
As can be seen easily, those values between ACCMM model with
$p_{_F}=0.3$ GeV and Isgur {\it et al.} model are
in large disagreement.
However, if we use $p_{_F}=0.5$ GeV, the result of the ACCMM model
becomes $1.03$, and  these two models are in a good
agreement for the value of $|V_{ub}/V_{cb}|$. 
\\

\noindent
{\bf \large 4. Conclusions}\\

We conclude  that
the value $p_{_F} \sim 0.3$ GeV, which has been commonly 
used in experimental analyses, has no theoretical or experimental
clear justification, even though there has been recently an assertion
that the prediction of heavy quark effective theory
approach~\cite{bigi}, far from the end-point region, 
gives approximately equal shape to the ACCMM model with
$p_{_F} \sim 0.3$ GeV.
Therefore, it is strongly recommended to determine the value of
$p_{_F}$ more reliably and independently, 
when we think of the importance of its role in experimental analyses.
It is particularly important in the determination of the value of
$|V_{ub}/V_{cb}|$. A better 
determination of $p_{_F}$ is also interesting theoretically since it
has its own physical correspondence related to the Fermi motion inside
$B$ meson.
In this context we 
calculated theoretically the value of $p_{_F}$ in the relativistic 
quark model using quantum mechanical variational method.
It turns out that $p_{_F}=0.50 \sim 0.54$ GeV, which is not far from
the value of $p_{_F}$ determined by comparing ACCMM model prediction
and the model independent lepton energy spectrum of ARGUS measurement,
$p_{_F}=0.27 \pm ^{0.22}_{0.27}$ GeV. 
The theoretically determined value of $p_{_F}$ is almost independent
of input parameters, $\alpha_s$, $m_b$, $m_c$, $m_{sp}$ and {\it etc.}
On the other hand, the experimentally determined value is strongly 
affected by the value of the input parameter $m_c$.

By using the same framework, we then calculated the ratio $f_B$/$f_D$, 
and obtaind the result which is larger, by 
the factor of about 1.3, than $\sqrt{M_D / M_B}$ given by the naive 
nonrelativistic analogy.
This result is in a pretty good agreement with the recent Lattice
calculations.
We also calculated the ratio $(M_{B^*}-M_{B})$/$(M_{D^*}-M_{D})$,
whose results suggest the subtlety of the mechanism which gives rise
to the value of this ratio.
\\

\noindent
{\em Acknowledgements}\\

\indent
The work  was supported 
in part by the Korean Science and Engineering  Foundation, 
Project No. 951-0207-008-2,
in part by Non-Directed-Research-Fund, Korea Research Foundation 1993,
in part by the Center for Theoretical Physics,
Seoul National University, 
in part by Yonsei University Faculty Research Grant,
in part by Dae Yang Academic Foundation,  and
in part by the Basic Science Research Institute Program,
Ministry of Education 1994,  Project No. BSRI-94-2425.

\pagebreak


\vspace*{3.5cm}
\centerline{{\large\bf Figure Captions}}
\vspace*{1cm}

\noindent
Fig. 1 The normalized lepton energy spectrum of
$B \rightarrow X_c l \nu$
for the whole region of electron energy from the recent
ARGUS \cite{argus2} measurement. Also shown are
the theoretical ACCMM model predictions, Eq. (\ref{f4}), using 
$p_{_F}=0,~0.27,~0.49$ GeV, corresponding to dashed-, full-,
dotted-line, respectively.
The minimum $\chi^2$  equals to 0.59 with $p_{_F} = 0.27$ GeV.
We fixed $m_{sp}=0.15$ GeV and $m_q=m_c=1.5$ GeV.


\begin{figure}[tb]
\vspace*{-2cm}
\centerline{\epsfig{figure=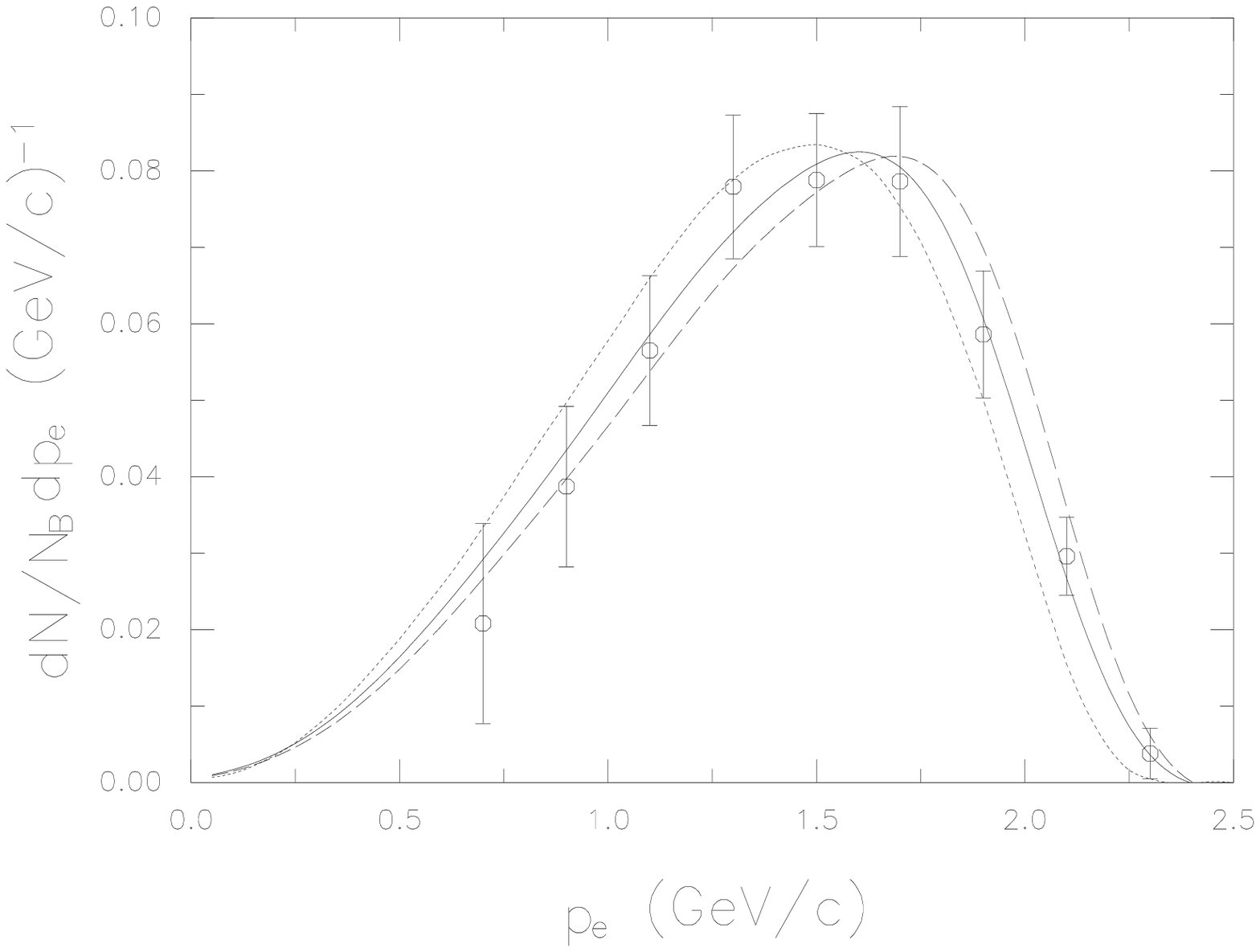}}

\end{figure}

\end{document}